\documentstyle[12pt,prd,aps,epsfig,preprint]{revtex}
\begin{document} \draft
\date{\today}
\title{QCD sum rule analysis of the coupling constants g$_{\rho\eta\gamma}$ and g$_{\omega\eta\gamma}$}

\author{C. Aydin~\thanks{coskun@risc01.ktu.edu.tr} and
        A. H. Yilmaz~\thanks{hakany@risc01.ktu.edu.tr}}
\address{ {\it Physics Department, Karadeniz Technical University,
Trabzon, Turkey}} \maketitle

\begin{abstract}
The coupling constants  g$_{\rho\eta\gamma}$ and
g$_{\omega\eta\gamma}$ are calculated using QCD sum rules method
by studying the three point ${\rho\eta\gamma}$ and
${\omega\eta\gamma}$ correlation functions. A comparison of the
results with the values of the coupling constants that are deduced
from the experimentally measured decay widths of $\rho\rightarrow
\eta\gamma$ and $\omega\rightarrow \eta\gamma$ decays is
performed.
\end{abstract}

\thispagestyle{empty} ~~~~\\ 

\newpage
\setcounter{page}{1}

The method of QCD sum rules is one of the efficients tools for
studying hadron physics. This method has been successfully applied
to calculate many hadronic observables, such as decay constants
and form factors \cite{R1,R2}. On the other hand, radiative
transitions between pseudoscalar (P) mesons have been an important
area of study in low-energy hadron physics for more than three
decades. These transitions have been analyzed within the
frameworks of phenomenological quark models, potential models, bag
models, and also by employing effective Lagrangian methods
\cite{R4,R5}. The radiative transitions $V\rightarrow P\gamma$ are
characterized by the coupling constants g$_{V\eta\gamma}$. Since
low energy hadron physics is governed by nonperturbative QCD, it
is very difficult to obtain the numerical values of these coupling
constants from the first principles. For this reason, some
specific nonperturbative methods have to be developed to be used
as calculational tools. Among these methods QCD sum rules have
proved to be very useful to extract the coupling constants. A
recent review of QCD sum rules method is provided in \cite{R6}
where more references can also be found.

In this work, we calculate the coupling constants
g$_{\rho\eta\gamma}$ and g$_{\omega\eta\gamma}$ associated with
the radiative decays $\rho\rightarrow\eta\gamma$ and
$\omega\rightarrow\eta\gamma$ by employing the traditional QCD sum
rules method which provides a model independent way to calculate
the coupling constants. The coupling constant g$_{\rho\eta\gamma}$
was previously calculated by T. M. Aliev et al. \cite{R7} in the
framework of light cone QCD sum rules. Our analysis, therefore,
complements the results obtained in that paper.

In accordance with the general strategy of QCD sum rules method,
we begin by considering the three point correlation function

\begin{equation}\label{e1}
  \Pi_{\mu\nu}(p,p^\prime)=\int d^{4}x d^4y e^{ip^\prime\cdot y}e^{-ip\cdot x}
  <o|T\{j_\mu^\gamma(0)j_{\nu}^V(x)j_{\eta}(y)\}|0>
\end{equation}
where the interpolating currents $j_{\nu}^V$ for vector meson
$\rho$ and $\omega$ are
$j_{\nu}^{\rho}=\frac{1}{\sqrt{2}}(\overline{u}\gamma_{\nu}u-\overline{d}\gamma_{\nu}d)$,
$j_{\nu}^{\omega}=\frac{1}{\sqrt{2}}(\overline{u}\gamma_{\nu}u+\overline{d}\gamma_{\nu}d)$,
respectively. We take $\eta-\eta^\prime$ mixing into account and
use the interpolating current for $\eta$ meson as
$j_{\eta}=\frac{1}{\sqrt{2}}
(\overline{u}i\gamma_{5}u+\overline{d}i\gamma_{5}d)cos\theta-(\overline{s}i\gamma_{5}s)sin\theta$
where $\theta$ is the  mixing angle in the quark-flavour basis.
The electromagnetic quark current is given as
$j_{\mu}^{\gamma}=e_u\overline{u}\gamma_{\mu}u+e_d\overline{d}\gamma_{\mu}d$,
where $e_u$ and $e_d$ denote the quark charges.

The theoretical part of the sum rule for the coupling constant
g$_{V\eta\gamma}$ is calculated by considering the perturbative
contribution and the power corrections from operators of different
dimensions to the three point correlation function. In the spirit
of QCD sum rules techniques, we consider the three point
correlation function in the Euclidian region defined by
$p^2=-Q^2\sim -1~~GeV^2$, $p'^2=-Q'^2\sim -1~~GeV^2$. In this
region, the perturbative contribution can be approximated by the
lowest order quark loop diagram shown in Fig. 1. Moreover, we
consider the power corrections from operators of different
dimensions, proportional to terms $<\overline{q}q>$,
$<\overline{q}\sigma\cdot Gq>$ and $<(\overline{q}q)^2>$. Since
the gluon condensate contribution proportional to $<G^2>$ is
estimated to be negligible for light quark systems, it is not
taken into account. We perform the calculations of the power
corrections in the fixed point gauge \cite{R8}. Moreover, we work
in the SU(2) flavour context with m$_u$=m$_d$=m$_q$ and we work in
the limit m$_q=0$. In this limit, the perturbative quark-loop
diagram does not make any contribution, and only contributions
result from the operators of dimensions d=3 and d=5 that are
proportional to $<\overline{q}q>$ and $<\overline{q}\sigma\cdot
Gq>$, respectively. The relevant Feynman diagrams for the
calculation of the power corrections are shown in Fig. 2 and Fig.
3.

We then calculate the three point correlation function
$\Pi_{\mu\nu}(p,p^\prime)$ using phenomenological considerations.
This function satisfies a double dispersion relation. We choose
the vector and pseudoscalar channels and by saturating this
dispersion relation by the lowest lying meson states in these
channels the physical part of the sum rule is obtained as
 \begin{equation}\label{e2}
  \Pi_{\mu\nu}(p,p^\prime)=\frac{<0|j_{\nu}^V|V>
  <V(p)|j_\mu^\gamma|\eta(p^\prime)><\eta|j_{\eta}|0>}
  {(p^2-m^2_V)({p^\prime}^2-m^2_\eta)}+...
\end{equation}
where the contributions from the higher states and the continuum
are shown by dots. The overlap amplitudes for vector and
pseudoscalar mesons are $<0|j_{\nu}^V|V>=\lambda_V u_V$ where
$u_V$ is the polarization vector of the vector meson $V=\rho,~
\omega$ and $<\eta|j_{\eta}|0>=\lambda_\eta$. The matrix element
of the electromagnetic current is given by
\begin{equation}\label{e3}
<V(p)|j_\mu^\gamma|\eta(p^\prime)>=
-i\frac{e}{m_V}g_{V\eta\gamma}K(q^2)\varepsilon^{\mu\nu\alpha\beta}p_\nu
u_\alpha q_\beta
\end{equation}
where $q=p-p^\prime$ and $K(q^2)$ is a form factor with K(0)=1.
This matrix element defines the coupling constant
g$_{V\eta\gamma}$ through the effective Lagrangian
\begin{equation}\label{e4}
{\cal
L}=\frac{e}{m_V}g_{V\eta\gamma}\varepsilon^{\mu\nu\alpha\beta}
\partial_\mu V_\nu\partial_\alpha A_\beta\eta
\end{equation}
describing the $V\eta\gamma$-vertex \cite{R9}.

After performing the double Borel transform with respect to the
variables $Q^2$ and ${Q^\prime}^2$, we obtain the sum rule for the
coupling constant  g$_{V\eta\gamma}$ in the form
\begin{eqnarray}\label{e5}
g_{V\eta\gamma}=&&\frac{m_V}{\lambda_V\lambda_\eta}
  e^{\frac{m_V^2}{M^2}}e^{\frac{m_\eta^2}{{M^\prime}^2}}
\left(e_u<\overline{u}u>\pm e_d<\overline{d}d>\right)
\nonumber \\
  &&
    \left (-\frac{3}{2}+\frac{5}{16}m_0^2\frac{1}{M^2}
    -\frac{3}{16}m_0^2\frac{1}{{M^\prime}^2}\right )\cos\theta
\end{eqnarray}
where the relation $<\overline{q}\sigma\cdot
Gq>=m_0^2<\overline{q}q>$ is used. In this expression the plus
sign is for $\rho$ meson and the minus sign is for $\omega$ meson.
In the numerical evaluation of the sum rule the values
$m_0^2=(0.8\pm 0.02)~~GeV^2$,
$<\overline{u}u>=<\overline{d}d>=(-0.014\pm 0.002)~~GeV^3$
\cite{R6}, and $m_\rho=0.77~~GeV$, $m_\omega=0.781~~GeV$,
$m_\eta=0.547~~GeV$ are used \cite{R10}. The overlap amplitudes
for vector meson states are calculated using the experimental
leptonic decay widths of $V\rightarrow e^+e^-$ decays \cite{R10}
and the values $\lambda_\rho=(0.17\pm 0.03)~~GeV^2$ and
$\lambda_\omega=(0.15\pm 0.02)~~GeV^2$ are obtained. The overlap
amplitude for $\eta$ meson state was determined earlier by QCD sum
rules analysis  as $\lambda_\eta=(0.23\pm 0.03)~~GeV^2$
\cite{R11}. We use the value of the mixing angle as
$\theta=-19^o\pm 2^o$ \cite{R11}.

The dependence of the coupling constants g$_{V\eta\gamma}$ on the
Borel parameters $M^2$ and ${M^\prime}^2$ are analyzed by studying
the independent variations of $M^2$ and ${M^\prime}^2$ in the
interval $0.6~~GeV^2\leq M^2,{M^\prime}^2\leq 1.4~~GeV^2$ since
these limits determine the allowed interval for the vector channel
\cite{R12}. We show the variation of the coupling constant
g$_{\rho\eta\gamma}$ and g$_{\omega\eta\gamma}$ as a function of
the Borel parameters $M^2$ for different values of ${M^\prime}^2$
in Fig. 4 and in Fig. 5, respectively. These figures indicate that
the sum rule is quite stable with these reasonable variations of
$M^2$ and ${M^\prime}^2$. We choose the middle value
$M^2=1~~GeV^2$ for the Borel parameter in its interval of
variation and obtain the coupling constants g$_{V\eta\gamma}$ as
g$_{\rho\eta\gamma}=1.2\pm 0.3$ and g$_{\omega\pi\gamma}=0.4\pm
0.06$ where the uncertainties result from the variations of $M^2$
and ${M^\prime}^2$ and from the estimated values of the vacuum
condensates.

If we use the effective Lagrangian given in Eq. 4, then the decay
width for $V\rightarrow\eta\gamma$ is obtained as
\begin{equation}\label{e6}
  \Gamma(V\rightarrow\eta\gamma)=\frac{\alpha}{24}
  \frac{(m_V^2-m_{\eta}^2)^3}{m_V^5}g_{V\eta\gamma}^2 ~~.
\end{equation}
We then utilize the measured decay widths
$\Gamma(\rho\rightarrow\eta\gamma)=(57\pm 10)$ keV and
$\Gamma(\omega\rightarrow\eta\gamma)=(5.5\pm 0.9)$ keV \cite{R9}
and obtain the coupling constants g$_{V\eta\gamma}$ as
g$_{\rho\eta\gamma}=1.42\pm 0.12$ and g$_{\omega\pi\gamma}=0.42\pm
0.03$. Our results are, therefore, in good agreement with the
coupling constants deduced from the experimental values of the
respective decay widths. Moreover, our result for
g$_{\rho\eta\gamma}$ is also consistent with the value
g$_{\rho\eta\gamma}=1.42\pm 0.2$ calculated by T. M. Aliev et al.
\cite{R7} in the framework of light cone sum rules, thus our study
employing traditional QCD sum rule method supplements the previous
light cone QCD sum rules calculation.

\begin{center}
{\bf ACKNOWLEDGMENT}
\end{center}

We like to thank  Profs. A. G\"{o}kalp and O. Y{\i}lmaz  for
suggesting this investigation to us and for helpful discussions
during the course of our work.


\newpage
\begin{figure}\vspace*{1.0cm}\hspace{4.5cm}
\epsfig{figure=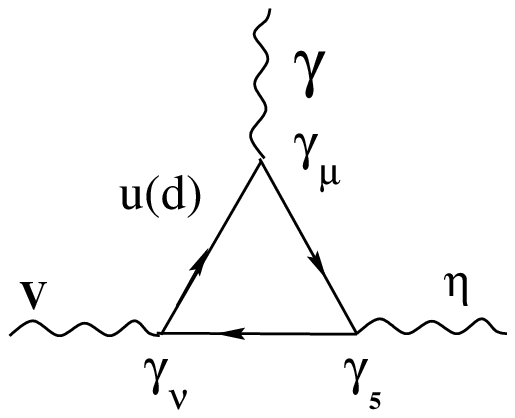,width=5cm,height=3cm,
angle=0}\vspace*{0.0cm} \caption{Quark loop diagram for
$V\eta\gamma$ vertex} \label{fig1}
\end{figure}

\begin{figure}
\vspace*{2.5cm}\hspace{1.5cm}
\epsfig{figure=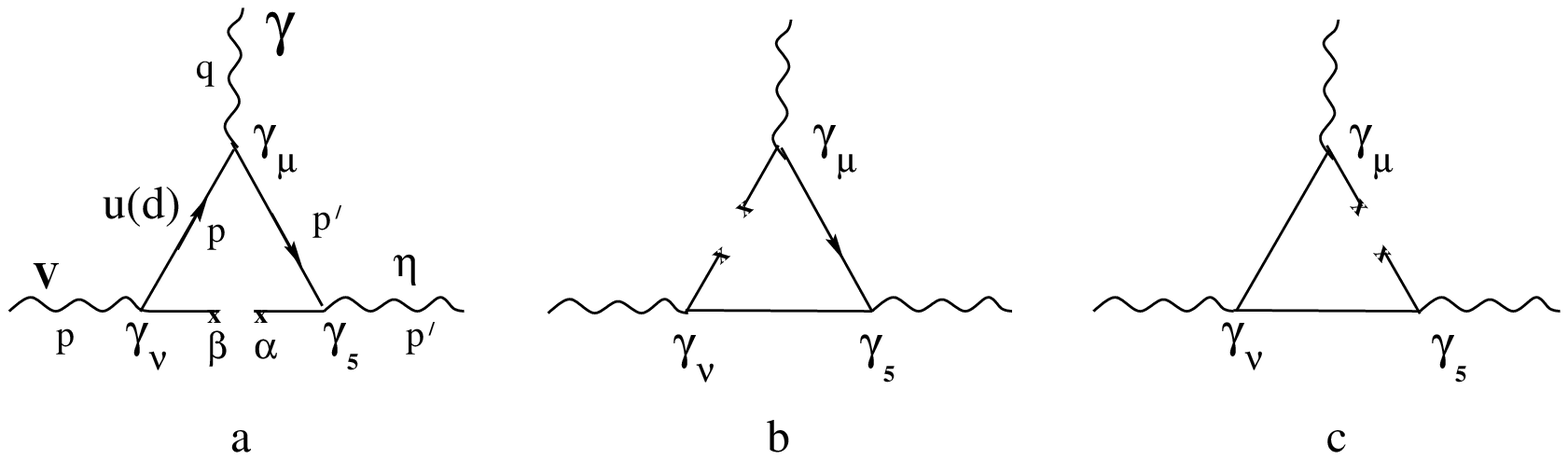,width=12cm,height=3cm,
angle=0}\vspace*{0.5cm} \caption{Operators of dimension 3
corrections proportional to $<(\overline{q}q)>$.} \label{fig2}
\end{figure}

\begin{figure}
\vspace*{2.5cm}\hspace{1.5cm}
\epsfig{figure=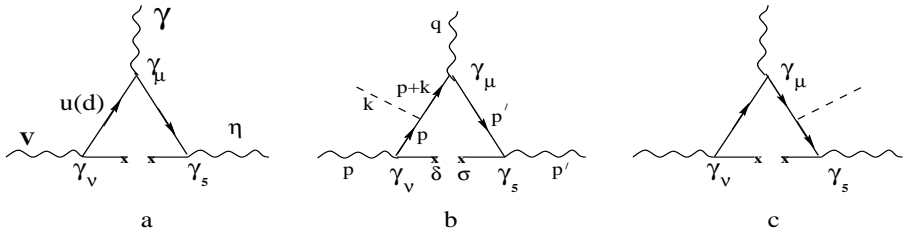,width=12cm,height=3cm}\vspace*{0.5cm}
\caption{Operators of dimension 5 corrections proportional to
$<\overline{q}\sigma\cdot Gq>$. The dot lines denote gluons.}
\label{fig3}
\end{figure}

\begin{figure}
\epsfig{figure=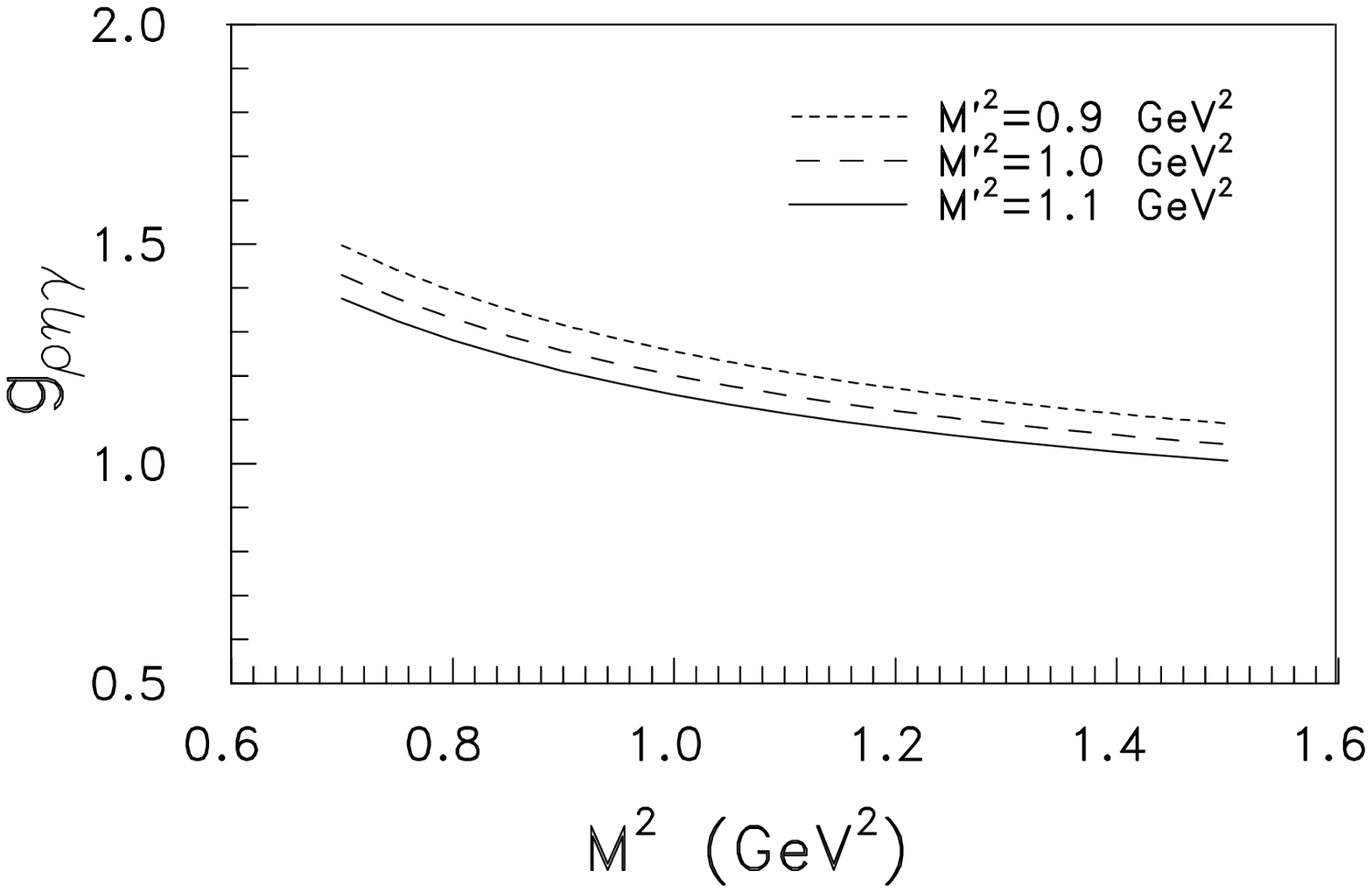,width=15cm,height=20cm} \vspace*{-3.0cm}
\caption{The coupling constant g$_{\rho\eta\gamma}$ as a function
of the Borel parameter $M^2$ for different values of
${M^\prime}^2$. } \label{fig4}
\end{figure}

\begin{figure}
\epsfig{figure=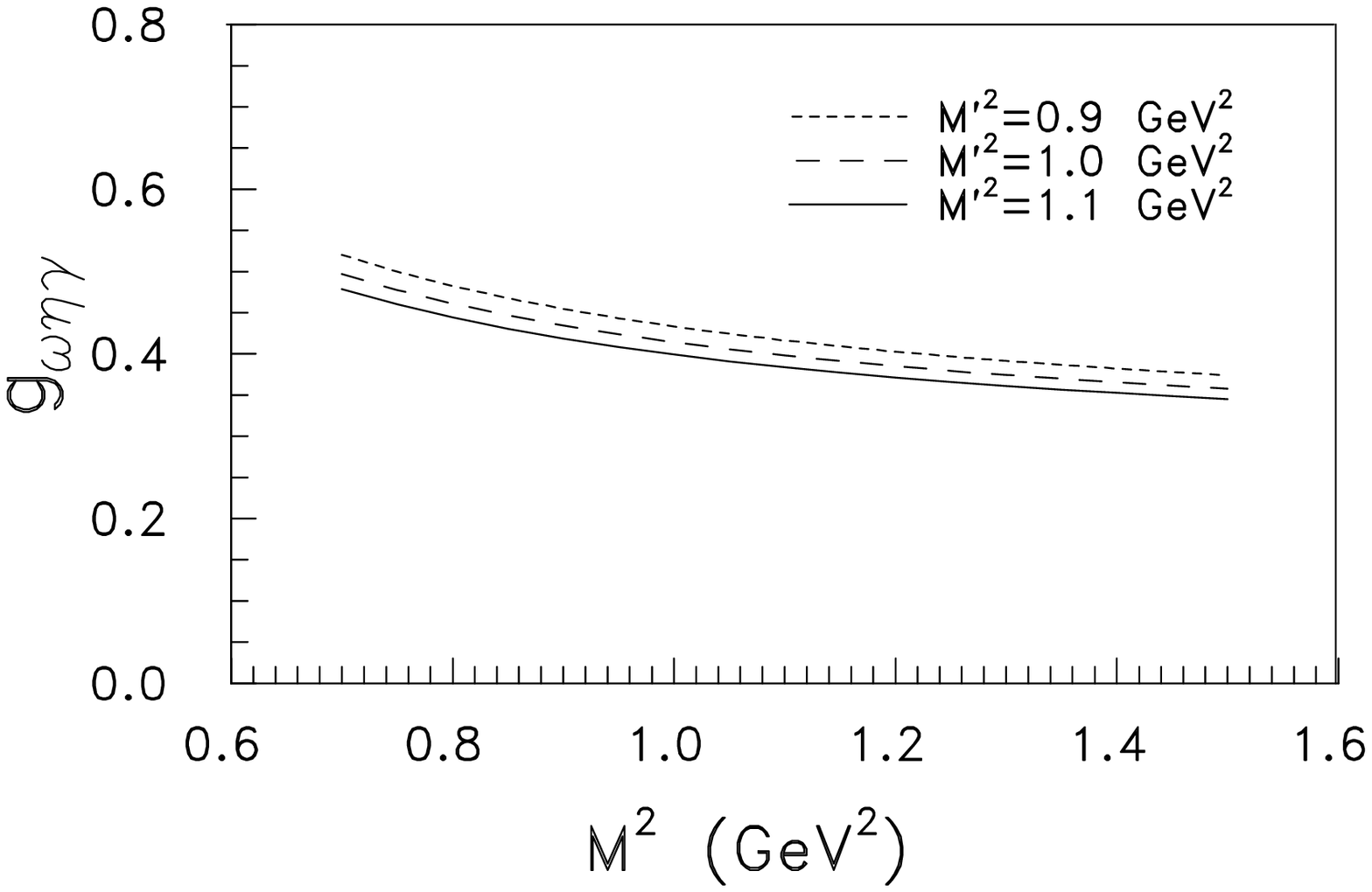,width=15cm,height=20cm} \vspace*{-3.0cm}
\caption{The coupling constant g$_{\omega\eta\gamma}$ as a
function of the Borel parameter $M^2$ for different values of
${M^\prime}^2$. } \label{fig5}
\end{figure}


\begin{references}
\bibitem{R1} M. A. Shifman, A. I. Vainstein, V. I. Zakharov, Nucl.
             Phys. {\bf B 147} (1979) 385 and 448.
\bibitem{R2} L. J. Reinders, S. Yazaki and H. R. Rubinstein, Nucl. Phys. {\bf B 196} (1985) 125.
\bibitem{R3} B.L. Ioffe, Nucl. Phys. {\bf B 188} (1981) 317; {\bf B 191}(1981) 591.

\bibitem{R4} A.Gokalp, O.Yilmaz, Eur. Phys. J. {\bf C 24}, (2002) 117
\bibitem{R5} P. J. O'Donnell, Rev. Mod. Phys. {\bf 53} (1981) 673
\bibitem{R6} P. Colangelo, A. Khodjamirian, in Boris Ioffe Festschrift, At the
Frontier of Particle Physics / Handbook of the QCD, edited by M.
Shifman World Scientific, Singapor.
\bibitem{R7}[6] T.M.Aliev, I.Kanik, A.Ozpineci, hep-ph/0212187
\bibitem{R8} A. V. Smilga, Sov. J. Nucl. Phys. {\bf 35} (1982) 271.
\bibitem{R9} A. I. Titov, T. -S. H. Lee, H. Toki, O. Streltsova, Phys. Rev.
              {\bf C 60} (1999) 035205.
\bibitem{R10} D. E. Groom et al., Eur. Phys. J. {\bf C15} (2000) 1;
\bibitem{R11} S-L. Zhu, W-Y. P. Hwang, Z-S. Yang, Phys. Lett.{\bf B 420} (1998) 8.
\bibitem{R12} V. L. Eletsky, B. L. Ioffe, Ya. I. Kogan, Phys. Lett.{\bf B 122} (1983) 423.

\end{references}
\end{document}